\documentclass[letterpaper]{article} 
\usepackage{aaai2026}  
\usepackage{times}  
\usepackage{helvet}  
\usepackage{courier}  
\usepackage[hyphens]{url}  
\usepackage{graphicx} 
\usepackage{natbib}  
\usepackage{caption} 
\usepackage{algorithm}
\usepackage{algorithmic}

\usepackage{url}  
\usepackage{booktabs}
\usepackage{amsfonts}  
\usepackage{nicefrac} 
\usepackage{microtype}   
\usepackage{fancyhdr}      
\usepackage{bm}
\usepackage{amsmath}
\usepackage{algorithmic}
\usepackage{algorithm}
\usepackage{multirow}
\usepackage{graphicx}
\usepackage{subfigure}
\usepackage{amsthm}
\usepackage{multicol}
\usepackage{pdfpages}
\usepackage{makecell}
\usepackage{xspace}
\usepackage{float}
\usepackage{threeparttable}
\usepackage{soul}
\usepackage{array}
\usepackage{dsfont}
\usepackage{pifont}
\usepackage{colortbl} 
\usepackage{xcolor}
\usepackage[most]{tcolorbox}
\usepackage{lipsum}
\usepackage{cuted} 
\definecolor{c1}{HTML}{d6ecf0}

\usepackage{etoolbox}
\makeatletter
\patchcmd{\@makecaption}
  {\parbox}
  {\advance\@tempdima-3em\parbox} 
  {}{}
\makeatother

\usepackage{newfloat}
\usepackage{listings}
\DeclareCaptionStyle{ruled}{labelfont=normalfont,labelsep=colon,strut=off} 
\floatstyle{ruled}
\newfloat{listing}{tb}{lst}{}
\floatname{listing}{Listing}
\title{Think Before You Segment: An Object-aware Reasoning Agent for \\ Referring Audio-Visual Segmentation}
\author{
    Jinxing Zhou\textsuperscript{\rm 1},
    Yanghao Zhou\textsuperscript{\rm 2},
    Mingfei Han\textsuperscript{\rm 1},
    Tong Wang\textsuperscript{\rm 1}, 
    Xiaojun Chang\textsuperscript{\rm 1,3},\\
    Hisham Cholakkal\textsuperscript{\rm 1},
    Rao Muhammad Anwer\textsuperscript{\rm 1}
    \\
}
\affiliations{
    \textsuperscript{\rm 1}Mohamed Bin Zayed University of Artificial Intelligence\\
    \textsuperscript{\rm 2}National University of Singapore\\
     \textsuperscript{\rm 3}University of Science and Technology of China\\
}

\usepackage{bibentry}

\begin{document}

\maketitle

\begin{abstract}

Referring Audio-Visual Segmentation (Ref-AVS) aims to segment target objects in audible videos based on given reference expressions.
Prior works typically rely on learning latent embeddings via multimodal fusion to prompt a tunable SAM/SAM2 decoder for segmentation, which requires strong pixel-level supervision and lacks interpretability.
From a novel perspective of explicit reference understanding, we propose TGS-Agent, which decomposes the task into a Think-Ground-Segment process, mimicking the human reasoning procedure by first identifying the referred object through multimodal analysis, followed by coarse-grained grounding and precise segmentation.
To this end, we first propose Ref-Thinker, a multimodal language model capable of reasoning over textual, visual, and auditory cues.
We construct an instruction-tuning dataset with explicit object-aware think-answer chains for Ref-Thinker fine-tuning.
The object description inferred by Ref-Thinker is used as an explicit prompt for Grounding-DINO and SAM2, which perform grounding and segmentation without relying on pixel-level supervision.
Additionally, we introduce R\textsuperscript{2}-AVSBench, a new benchmark with linguistically diverse and reasoning-intensive references for better evaluating model generalization.
Our approach achieves state-of-the-art results on both standard Ref-AVSBench and proposed R\textsuperscript{2}-AVSBench.
Code will be available at \url{https://github.com/jasongief/TGS-Agent}.
\end{abstract}


\section{Introduction}\label{sec:intro}

\begin{figure}[!t]
\centering
 \includegraphics[width=1\columnwidth]{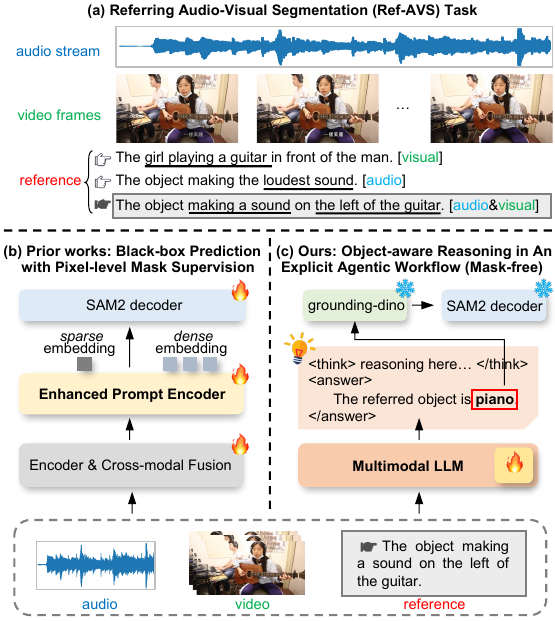}
\caption{(a) Illustration of Ref-AVS task. (b) Prior works focus on enhancing sparse and dense prompt embeddings via implicit transformer fusion, requiring strong pixel-level supervision for model training. (c) Our method first performs explicit reasoning over the reference using a MLLM to identify the referred object, and then generates its bounding box and segmentation mask within an agentic workflow.
 }
 \label{Fig:intro}
\end{figure}

In recent years, the trend in artificial intelligence research has been shifting toward omni-modal understanding, where models are expected to jointly perceive and reason across multiple modalities.
Among various topics, the Referring Audio-Visual Segmentation (Ref-AVS) task aims to segment target objects in audible video scenes based on given natural language expressions (a.k.a \textit{reference}).
As shown in Fig.~\ref{Fig:intro}(a), the reference may contain either single-modal or multi-modal cues, requiring the model to automatically analyze and integrate relevant cues from audio, visual, and textual modalities.
For instance, the reference ``\textit{The object making a sound on the left of the guitar}'' demands integrating textual semantics, spatial visual cues, and temporal audio patterns to segment the correct object `piano' across video frames.

Research on the Ref-AVS task remains an active and evolving area.
The pioneering work, EEMC~\cite{wang2024ref}, employs multiple transformer blocks to model interactions among audio, visual, and textual modalities.
The resulting integrated multimodal features are then used as cues to prompt a segmentation decoder, Mask2Former~\cite{cheng2021mask2former}, to generate binary masks of the referred object.
Recent approaches leverage more powerful segmentation models such as SAM~\cite{kirillov2023sam} and SAM2~\cite{ravi2024sam2}, extending them to handle multimodal signals.
As shown in Fig.~\ref{Fig:intro}(b), a common focus of these methods is to enhance the prompt encoder of SAM or SAM2 (\textit{i.e.}, the sparse and dense prompt embeddings) by modeling more informative multimodal cues.
For instance, TSAM~\cite{radman2025tsam} utilizes cross-attention mechanism to fuse audio-visual-text modalities. Another recent method, SAM2-LOVE~\cite{wang2025sam2love}, proposes to summarize multimodal features into a learnable \texttt{[seg]} token, which serves as an improved sparse prompt for SAM2.
The mask decoder of SAM/SAM2 is also fine-tuned to better adapt to the Ref-AVS task.
Although these methods achieve strong performance, they rely on pixel-level ground truth masks for supervision, and their segmentation processes function as black boxes, lacking interpretability.
In contrast, our work explores a ground truth \textbf{\textit{mask-free}} and \textbf{\textit{more explainable}} approach.

Let's reconsider how humans naturally approach the Ref-AVS task.
Given a reference, \textit{e.g.}, ``\textit{The object making a sound on the left of the guitar}'' shown in Fig.~\ref{Fig:intro}(a), we first analyze this reference text, observe video frames to locate the \textit{guitar}, shift our attention to its \textit{left region}, and listen to the audio to identify the object which \textit{is making a sound}.
Following this chain of reasoning, we can clearly determine the referred object, \textit{i.e.,} the \textit{piano}.
Once the \textit{piano} is identified, we can ground its position and precisely segment the corresponding pixels across video frames.
We refer to this procedure as a `\textbf{\textit{Think-Ground-Segment (TGS)}}' decision process.
This critical step is often overlooked by prior methods: before performing fine-grained segmentation, humans can first explicitly identify the target object by thinking about multimodal (audio, visual, textual) inputs.

Motivated by this observation, we propose \textbf{TGS-Agent}, an agentic framework with an explicit, explainable, object-aware reasoning chain for the Ref-AVS task (illustrated in Fig.~\ref{Fig:intro}(c)).
Thanks to the advancements in foundation models for object detection and segmentation, the \textit{Ground} and \textit{Segment} steps can be effectively handled. 
In our framework, we adopt Grounding-DINO~\cite{liu2024gdino} and SAM2~\cite{ravi2024sam2} as the tools for these two stages, respectively.
Given an appropriate textual prompt (related to target objects in Ref-AVS setting), Grounding-DINO generates its corresponding bounding box (object text $\rightarrow$ bbox).
Subsequently, SAM2 takes the bbox as prompt and accurately produces the corresponding pixel mask (bbox $\rightarrow$ mask).
Therefore, a critical challenge is how to address the \textit{Think} step: identify the referred object according to multimodal signals (reference $\rightarrow$ object text).
Since the output of the \textit{Think} step is an open-ended textual description related to the referred object, we introduce a multimodal large language model (MLLM), referred to as the \textbf{Ref-Thinker}, to infer the target object from the multimodal context (\textit{i.e.}, the reference alongside its audio and visual counterparts).
To enhance the reasoning and instruction-following capabilities of Ref-Thinker, we utilize Gemini-1.5-Pro with carefully designed prompts to construct an instruction-tuning set, containing explicit think-answer reasoning chains. More details will be introduced in Sec.~\ref{sec:ref_thinker}.
As a result, the fine-tuned MLLM is able to reason over the audio, visual, and reference inputs and generate reliable descriptions of the referred object.
We explore the impact of using simplified (category only) and fine-grained (category with more details like its attribute or spatial location) description texts (See more discussion in Sec.~\ref{sec:ablations}).

We evaluate TGS-Agent on the existing Ref-AVSBench dataset~\cite{wang2024ref}. In addition, we propose 
\textbf{R\textsuperscript{2}-AVSBench}, a new evaluation set designed to be more reasoning-intensive.
Specifically, we observe that references in a portion of the Ref-AVSBench test set tend to be straightforward.
For example, the target object name directly appears in the reference (\textit{e.g.}, `\textit{The clarinet being played by a man}').
Moreover, Ref-AVSBench references are constructed using fixed templates, limiting their linguistic diversity.
To improve this, we propose R\textsuperscript{2}-AVSBench, which features references with greater lexical and structural diversity, and which require deeper reasoning to interpret.
We prompt Gemini-1.5-Pro to achieve the reference transformation.
Further details will be provided in Sec.~\ref{sec:bench}.
Our R\textsuperscript{2}-AVSBench can serve as a more effective benchmark for evaluating a model's generalizability across diverse reference types.

In summary, our main contributions are:
\textbf{(1)} We propose TGS-Agent, decoupling the Ref-AVS task into a `{Think-Ground-Segment}' agentic workflow. TGS-Agent provides a new paradigm that is mask supervision-free and also more explainable.
\textbf{(2)} We propose Ref-Thinker, a MLLM with enhanced object-aware reasoning ability, which generates explicit descriptions of the referred object by analyzing multimodal signals. 
An instruction tuning set is constructed to support Ref-Thinker training.
\textbf{(3)} We propose R\textsuperscript{2}-AVSBench, a new evaluation benchmark featuring more challenging and reasoning-intensive references. It serves as an additional testbed to evaluate model performance in cross-reference scenarios.
\textbf{(4)} Our method achieves state-of-the-art results on both Ref-AVSBench and R\textsuperscript{2}-AVSBench, significantly outperforming existing methods.

\section{Related Work}~\label{sec:related_work}

\noindent\textbf{Audio-Visual Scene Understanding} aims to analyze the rich and dynamic audio-visual signals present in real-world environments, which may facilitate compelling applications in areas such as video generation~\cite{google2025veo,mao2025autoregressive,mao2024tavgbench}, conversational assistant~\cite{xu2025qwen2}, and virtual reality.
In academia, researchers have investigated various fundamental tasks, such as audio-visual event parsing~\cite{tian2018audio,tian2020unified,liu2025towards,zhou2021psp,zhou2022cpsp,zhou2023improving,zhou2024label,zhou2024advancing,zhao2025multimodal,zhou2025dense,zhou2025towards,yu2025prefm}, sound source localization~\cite{zhao2018sound,chen2021localizing,park2024can,um2025object}, and audio-visual caption or question answering~\cite{shen2023fine,li2022learning,yun2021pano,yang2022avqa,wu2025avqacl,li2024object,li2025patch}.
These tasks involve audiovisual comprehension across diverse inputs, ranging from basic audio-image pairs and short video clips to complex, untrimmed long-form audible videos.
The studied referring audio-visual segmentation task is also part of this field, requiring the understanding of audio, visual, and textual modalities.

\noindent\textbf{Audio-Visual Segmentation} aims to generate the pixel-level segmentation maps of the sounding objects present in audible videos~\cite{zhou2022avs,zhou2023avss,guo2025audio}.
The baseline AVS model~\cite{zhou2022avs} adopts a convolution-based decoder.
Subsequent studies explore various transformer-based~\cite{gao2024avsegformer,li2023catr,ling2024transavs,zhoualoha,guo2024enhance,ma2024stepping,yang2024cooperation,zhou2025mettle}, diffusion-based~\cite{mao2025contrastive}, and SAM-based methods~\cite{mo2023av,liu2024annotation,wang2024prompting,bhosale2025unsupervised,luo2025tavis}.
Among them, AL-Ref-SAM2\cite{huang2025unleashing} 
is the most relevant to our work.
AL-Ref-SAM2 performs sample-specific inference at test time using GPT-4, requiring carefully crafted prompts to select pivot frames and describe candidate bounding boxes for segmentation.
Since GPT-4 cannot directly process audio, the audio stream is converted into a textual description, where a pre-trained audio classification model~\cite{chen2022beats} is used to predict the audio categories.
However, the accuracy of these audio categories is highly dependent on the pre-trained classifier and limited by its predefined, closed-set label vocabulary.
Moreover, AL-Ref-SAM2 is not well-suited for Ref-AVS task, where: 1) the referred object in the text may be related to, but not necessarily, the actual sound source, making direct reliance on audio categories potentially misleading; and 2) the diverse and complex references in Ref-AVS require a unified understanding of audio, visual, and textual modalities.
In contrast, our approach fine-tunes an open-source LLM with enhanced reasoning capabilities over the reference expression, enabling it to accurately identify the true target object across video frames and simplify segmentation mask prediction.

\noindent\textbf{Referring Audio-Visual Segmentation}
aims to generate binary masks of the object referred to by a natural language expression involving either or both auditory and visual cues.
The baseline method, EEMC~\cite{wang2024ref}, integrates audio, visual, and textual features through a series of transformers to form a unified multimodal cue, which then prompts a segmentation decoder, Mask2Former~\cite{cheng2021mask2former}, via cross-attention to accurately identify and delineate the object of interest.
The recent method TSAM~\cite{radman2025tsam} enhances SAM~\cite{kirillov2023sam} encoder with a temporal modeling branch and designs data-driven multimodal sparse and dense prompts for SAM's decoder. 
In particular, reference texts are used to guide the selection of relevant audio cues, which then interact with visual cues to form \textit{sparse} prompts.
The \textit{dense} prompts are also generated through cross-attention between visual-audio and visual-text features.
SAM2-LOVE~\cite{wang2025sam2love} leverages SAM2~\cite{ravi2024sam2} for Ref-AVS task. 
It integrates textual, audio, and visual representations into a learnable segmentation token \texttt{[seg]}.
To enhance spatial-temporal consistency across video frames, SAM2-LOVE employs token propagation and token accumulation strategies to strengthen \texttt{[seg]}, which serves as a more effective \textit{sparse} prompt for SAM2's decoder. 
Unlike prior methods, we utilize a reasoning-enhanced MLLM to \textit{explicitly} identify the referred object. 
Our method is more explainable and does not require pixel-level supervision.

\begin{figure}[!t]
\centering
 \includegraphics[width=1\columnwidth]{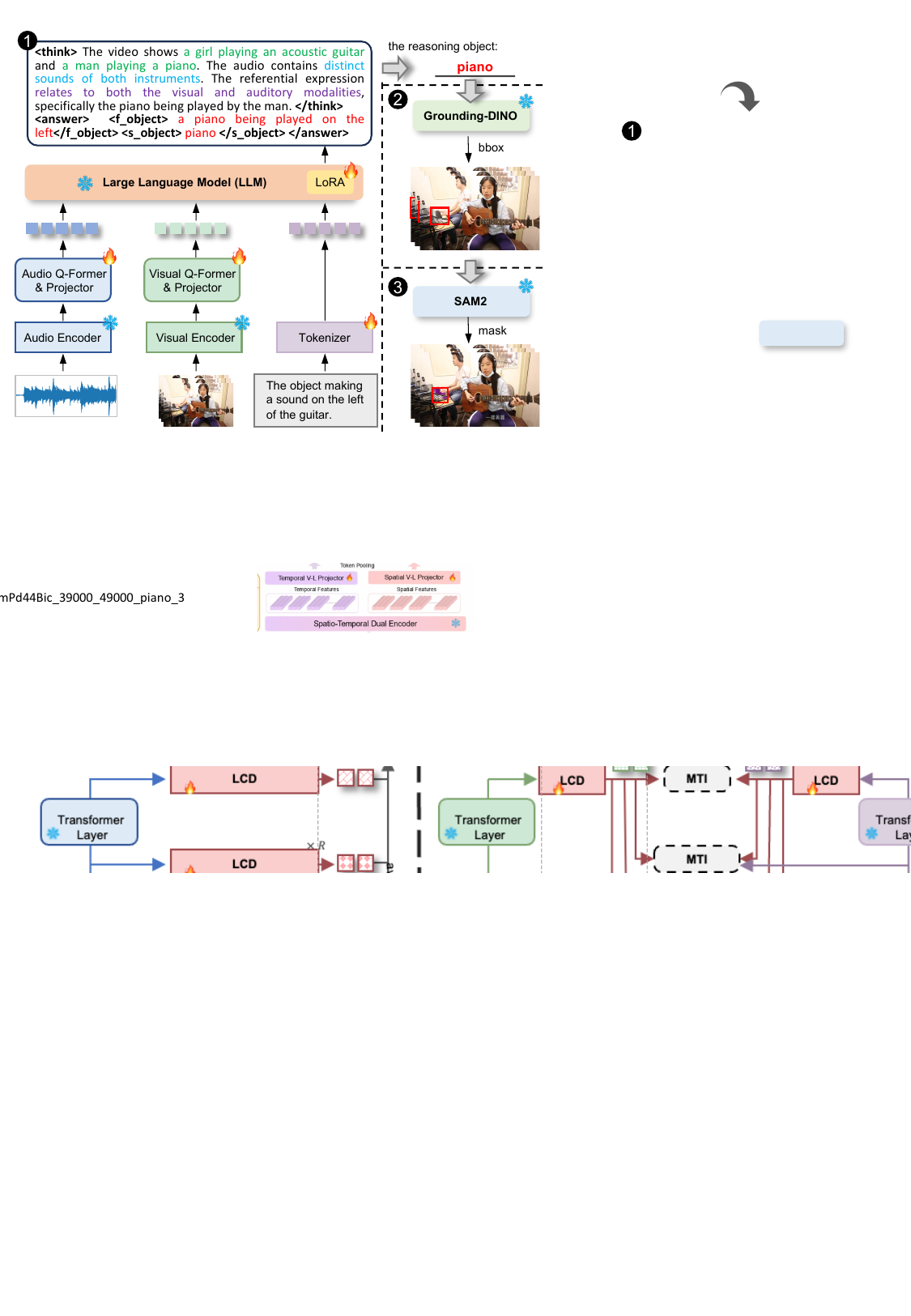}
\caption{Workflow of our Think-Ground-Segment Agent.
 }
 \label{Fig:framework}
\end{figure}

\section{Our Method}~\label{sec:method}

In this section, we first introduce the proposed TGS-Agent framework for the Ref-AVS task (Sec.~\ref{sec:TGS_agent}).
Next, we present Ref-Thinker, the core component of TGS-Agent, which is a reasoning-enhanced MLLM designed to explicitly generate textual descriptions of the referred object.
We describe its architecture and training strategy in Sec.\ref{sec:ref_thinker}.

\subsection{TGS-Agent: An Object-aware Ref-AVS Agent}\label{sec:TGS_agent}

As discussed in the Introduction, our TGS-Agent decomposes the Ref-AVS task into multiple object-aware processing steps.
The overall system flow is shown in Fig.~\ref{Fig:framework}.
Below, we elaborate on each stage along with its corresponding agentic tool.

\noindent\textbf{Think}. As a core innovation of our method, this first step aims to clearly answer what the referred object is.
To achieve this, we propose Ref-Thinker, a reasoning-enhanced MLLM, which can digest multimodal inputs (\textit{i.e.}, audio, visual, and reference text) and output a description related to the referred object with an explicit reasoning chain.
Specifically, the output text follows the format below:
\begin{tcolorbox}[
  colback=gray!10,      
  colframe=black,       
  arc=3mm,              
  boxrule=0.8pt,        
  left=4pt, right=4pt,  
  top=4pt, bottom=4pt,  
  fonttitle=\bfseries
]
\textbf{== Object-aware Reasoning Chain of Ref-Thinker ==} \\
\textbf{$<$think$>$}\\
\hspace*{3mm}The referential expression is ``xxx''.\\
\hspace*{3mm}The video shows xxx (video analysis here).\\
\hspace*{3mm}The audio contains xxx (audio analysis here).\\
\hspace*{3mm}The reference related to xxx (modality analysis here).\\
\textbf{$<$/think$>$}\\
\textbf{$<$answer$>$}\\
\hspace*{3mm}\textbf{$<$f\_object$>$}\\
\hspace*{6mm} A fine-grained description of the referred object, including detailed attributes such as color, shape, or spatial location (\textit{e.g.}, \textit{a guitar being played on the right}). \\
\hspace*{3mm}\textbf{$<$/f\_object$>$}\\
\hspace*{3mm}\textbf{$<$s\_object$>$}\\
\hspace*{6mm} A simplified description of the referred object, specifying only its category name (\textit{e.g.}, \textit{guitar}). \\
\hspace*{3mm}\textbf{$<$/s\_object$>$}\\
\textbf{$<$/answer$>$}
\end{tcolorbox}

\noindent
In the thinking process, Ref-Thinker confirms the reference expression, analyzes multimodal contents, and decides which modality should be focused.
Based on these analyses, Ref-Thinker provides answers with explicit information about the referred object.
Notably, we explore two types of object descriptions: \textit{fine-grained} (`f\_object') and \textit{simplified} (`s\_object').
The former one includes more details, such as the object's appearance, attributes, or spatial relations (while remaining concise, typically fewer than 10 words), whereas the latter simplified description contains only the object category.
We make this design considering that both the phases and class names are frequently used as prompts in object detection and segmentation. 
{Moreover, the fine-grained description may be more beneficial for video scenes that contain different object instances that have a similar appearance or the same category (as shown in Fig.~\ref{Fig:f_better_than_s}).}

The above process can be formalized as: 
\begin{equation}
 \texttt{\textbf{Think}}(\bm{A}, \bm{V}, \bm{R}, \bm{P}) \rightarrow \bm{T},
\end{equation}
where $\bm{A}$, $\bm{V}$, and $\bm{R}$ denote the audio stream, video frames, and reference, respectively.
$\bm{P}$ is the user prompt text used to guide the Ref-Thinker (detailed in Sec.~\ref{sec:ref_thinker}).
$\bm{T}$ represents the generated reasoning text description.
For convenience, we denote the fine-grained and simplified description of the referred object in $\bm{T}$ as $\bm{T_f}$ and $\bm{T_s}$, respectively.

\noindent\textbf{Ground}.
After obtaining the explicit object description $\bm{T_f}$ or $\bm{T_s}$, the second step is to generate a bounding box (bbox) for the target object in each video frame.
Regardless of whether the given reference is grounded in audio, visual, or both modalities, as illustrated in Fig.~\ref{Fig:intro}(a), it ultimately corresponds to a specific visible object (when it is present in the frame).
We adopt Grounding-DINO\cite{liu2024gdino} as the tool for bounding box generation.
Grounding-DINO is an advanced object detection model that combines the Transformer-based detector DINO\cite{zhang2022dino} with large-scale grounded pre-training, enabling it to detect arbitrary objects based on human language inputs, such as category names or referring expressions.
Therefore, our object descriptions ($\bm{T_f}$/$\bm{T_s}$) are closely aligned with the model's input format and intended usage.
We denote this process as:
\begin{equation}
    \texttt{\textbf{Ground}}(\bm{T_f}/\bm{T_s}, \bm{V}) \rightarrow \bm{B},
\end{equation}
where $\bm{V}=\{v_i\}_{i=1}^N$ denotes the $N$ video frames, and $\bm{B}=\{(x_1,y_1),(x_2,y_2)\}^N$ is corresponding bounding box set, with $(x_1, y_1)$ and $(x_2, y_2)$ indicating the coordinates of the top-left and bottom-right corners, respectively.
Notably, two key hyper-parameters are involved in the bbox generation: $\tau_\text{bbox}$ controls the minimum confidence score for a bounding box to be considered valid, while the $\tau_\text{text}$ sets the minimum confidence score for the text to be matched or recognized.

\noindent\textbf{Segment.}
After obtaining the bounding box $\bm{B}$ of the referred object in each video frame, we use this explicit and strong guidance as a sparse prompt to drive the powerful segmentation foundation model, SAM2~\cite{ravi2024sam2}, to generate the corresponding segmentation masks.
In particular, if the bounding box does not exist (\textit{i.e.}, no matching object is found in the video frame), the segmentation mask defaults to all background pixels.
We formalize this process as:
\begin{equation}
    \texttt{\textbf{Segment}}(\bm{B}, \bm{V}) \rightarrow \bm{M}.
\end{equation}
$\bm{M}=\{m_i\}_{i=1}^N$ is the set of binary masks for $N$ video frames.
Notably, the prior SOTA method~\cite{wang2025sam2love} also employs SAM2 but requires fine-tuning its mask decoder. In contrast, our method leverages the frozen SAM2 and achieves superior performance.

In this way, our TGS-Agent completes the transformation from audiovisual streams + reference text $\rightarrow$ object description $\rightarrow$ bbox $\rightarrow$ mask, demonstrating an object-aware, reliable, and explainable decision process for Ref-AVS task.

\subsection{Ref-Thinker: A Reasoning-enhanced MLLM}\label{sec:ref_thinker}

\noindent\textbf{Architecture.} 
As shown on the left side of Fig.~\ref{Fig:framework}, our Ref-Thinker is implemented as an audio-visual LLM. 
Given temporal audio and visual streams, modality-specific encoders are used to extract segment-level features.
The resulting audio and visual features are then compressed into a smaller number of token embeddings using independent Q-Formers~\cite{li2023blip}.
A projector module, implemented as two MLP layers, is further applied to align the audio/visual feature with the textual features processed by the LLM.
The textual input consists of two parts: a task-relevant user prompt containing the specific reference expression, and the corresponding reasoning chain as the expected output.
We provide the template of the user prompt ($\bm{P}$) below:
\begin{tcolorbox}[
  colback=gray!10,      
  colframe=black,       
  arc=3mm,              
  boxrule=0.8pt,        
  left=4pt, right=4pt,  
  top=4pt, bottom=4pt,  
  fonttitle=\bfseries
]
\textbf{===== User Prompt Template for Ref-Thinker =====} \\
This is a video:$<$video\_start$>$$<$video$>$$<$video\_end$>$. This is an audio:$<$audio\_start$>$$<$audio$>$$<$audio\_end$>$.
Given the referential expression xxx, analyze the video and audio, and then generate a reasoning chain ($<$think$>$) and final answer ($<$answer$>$).
Your output must follow this format: (reasoning chain introduced in Sec.~\ref{sec:TGS_agent}).
\end{tcolorbox}
\noindent Audio and visual feature embeddings are used to replace the placeholders $<$audio$>$ and $<$video$>$ tokens in the user prompt. The resulting textual input is then tokenized and passed to the LLM backbone for text generation.

\noindent\textbf{Training Recipe.}
We train our Ref-Thinker in two phases.
During the \textit{pretraining} phase, the LLM is frozen, and domain-specific caption datasets\cite{kim2019audiocaps,lin2023video} are independently used to train the audio and visual Q-Formers along with their corresponding projectors.
Next, in the \textit{instruction tuning} phase, we apply the LoRA~\cite{hu2022lora} technique for parameter-efficient tuning of the LLM.
This phase not only aims to improve the model's instruction following ability, but, more importantly, enables the MLLM to better learn an object-aware reasoning chain, as introduced in Sec.~\ref{sec:TGS_agent}.
To support this, we construct an instruction tuning set.
Specifically, given a reference expression and a video from the training set of the official Ref-AVSBench dataset~\cite{wang2024ref}, we leverage Gemini-1.5-Pro~\cite{google2024gemini} with carefully designed prompts to analyze the video and generate an object-aware reasoning chain that strictly follows the think-answer format shown in Sec.~\ref{sec:TGS_agent}.
Detailed prompt is provided in our Appendix A.
Supervised with this high-quality instruction tuning set, our Ref-Thinker can effectively think about the reference (and audiovisual signals) and generate accurate object descriptions with an explicit reasoning chain.
For both training phases, the autoregressive cross-entropy loss is used for optimization.

\section{ {R\textsuperscript{2}-AVSBench} }~\label{sec:bench}

We propose R\textsuperscript{2}-AVSBench as an additional \textbf{R}easoning-enhanced evaluation set for the \textbf{R}ef-AVS task, motivated by two key factors:
1) We observed that some reference expressions in the standard Ref-AVSBench~\cite{wang2024ref} test set are relatively simple.
For example, `\textit{The couch sit by a woman}' (referred object: couch) and `\textit{The object making a sound by using a saw}' (referred object: man holding the saw).
In these cases, the object name \textit{couch} or key cues \textit{saw} are explicitly mentioned in the reference, which may lead to shortcut learning by the model without fully reasoning over the reference context.
Additionally, references in Ref-AVSBench are constructed using fixed, manually predefined templates, which limits their linguistic diversity.
2) Since our work emphasizes object-aware reasoning over the \textit{reference}, we are interested in evaluating the model's ability to handle more challenging and varied reference types. 

\begin{figure}[!t]
\centering
 \includegraphics[width=1\columnwidth]{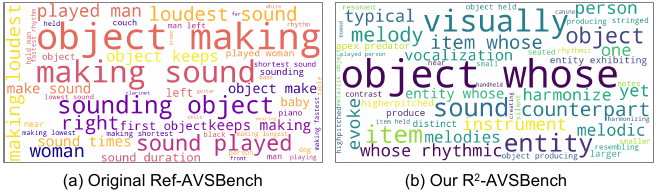}
\caption{Word clouds of different reference types. More detailed reference examples are provided in Appendix C.
 }
 \label{Fig:word_cloud}
\end{figure}

To this end, the references in our R\textsuperscript{2}-AVSBench are designed with greater lexical and structural diversity, which also requires deeper reasoning.
As illustrated by the word cloud visualization in Fig.~\ref{Fig:word_cloud}, the new references replace many explicit object names like `man' with abstract terms such as `counterpart', and use more relative pronouns like `whose' or `whose rhythmic'.
For instance, the aforementioned two reference examples are transformed into `\textit{The item visually serving as a shared seating platform for the audio discourse}' and `\textit{The entity rhythmically wielding a tool known for forestry tasks in earlier eras}'.
Both transformed references avoid directly revealing the target object, which also requires more reasoning about the object function, commonsense knowledge, audio information, and other contextual cues. Notably, we ensure that the new references target the same objects as those in the original Ref-AVSBench references. This design allows us to reuse the pixel-level masks provided by Ref-AVSBench for evaluating predictions and comparing model performance between the original and transformed references.
Additional reference examples from our R\textsuperscript{2}-AVSBench, along with comparisons to those in Ref-AVSBench, are provided in Appendix C.

We perform the reference transformation with the aid of Gemini-1.5-Pro, and the corresponding prompt is provided in Appendix B.
From the test set of Ref-AVSBench, we select a total of 400 unique videos for inclusion in our R\textsuperscript{2}-AVSBench, each paired with a more challenging reference.
Ref-AVSBench contains test subsets based on seen and unseen object categories, and we preserve this division in our R\textsuperscript{2}-AVSBench, resulting in 220 and 180 test videos for seen and unseen sets, respectively.
We also analyze the average number of words per reference, which is 11.73 for R\textsuperscript{2}-AVSBench and 7.08 for Ref-AVSBench, highlighting the increased linguistic diversity and complexity of our benchmark.
To ensure the high quality of R\textsuperscript{2}-AVSBench, each generated reference is further verified by human annotators.
Specifically, the references produced by Gemini, along with their corresponding audio and video, are reviewed by humans.
If a reference contains hallucinations, factual errors, requires minimal reasoning, or does not correspond to the original target object, it is revised or improved by the human annotators.

\section{Experiments}\label{sec:Exp}
\subsection{Experimental Setup}
\noindent\textbf{Datasets.}
We evaluate our method on both the Ref-AVSBench dataset~\cite{wang2024ref} and our proposed R\textsuperscript{2}-AVSBench. 
Ref-AVSBench contains 4,000 10-second videos and 20,000 manually annotated expressions, covering 51 object classes.
The training and validation sets contain 2,908 and 276 videos, respectively. 
The test set is divided into three subsets: 292 videos in the \textit{Seen} split, where object classes appear in the training set; 269 videos in the \textit{Unseen} split, which includes 13 additional novel categories not seen during training; a special \textit{Null} split, in which the expression refers to an object that does not exist in video frames.
For evaluation on R\textsuperscript{2}-AVSBench, we directly test the performance of models trained on Ref-AVSBench.

\noindent\textbf{Evaluation Metrics.}
Following prior works~\cite{wang2024ref,wang2025sam2love}, we adopt the Jaccard index $\mathcal{J}$ and F-score $\mathcal{F}$ as primary evaluation metrics. For the \textit{Null} set, a metric $\mathcal{S}$ is used by computing the ratio between the predicted mask area and the background area. A lower $\mathcal{S}$ indicates fewer false-positive pixels are predicted.

\noindent\textbf{Implementation Details.}
For the Ref-Thinker, we use CLIP-ViT-L/14~\cite{radford2021learning} and BEATs~\cite{chen2022beats} as the visual and audio encoders, respectively.
The number of learnable query tokens is set to 32.
We adopt LLaMA-2-7b-chat as the base LLM.
LoRA is applied with a rank 8 and a scaling factor 16.
The batch size is 4, and the LLM is fine-tuned for 6 epochs.
We use AdamW optimizer with an initial learning rate of 1e-4. 
We train our model on four NVIDIA A100-SXM4-40GB GPUs using bf16 precision.
We employ the Swin-T-based Grounding-DINO for object detection and the Hiera-Large-based SAM2 for segmentation, both with frozen parameters.

\begin{table*}[t]
\centering
\resizebox{1\textwidth}{!}{
\begin{tabular}{l|c|c|ccc|ccc|ccc|c}
\toprule
& & &
\multicolumn{3}{c|}{\textbf{Seen (S)} $\uparrow$} &
\multicolumn{3}{c|}{\textbf{Unseen (U)} $\uparrow$} &
\multicolumn{3}{c|}{\textbf{Mix (S+U)} $\uparrow$} &
\textbf{Null} $\downarrow$ \\
\textbf{Method} & \textbf{Venue} & \textbf{Task} &
$\mathcal{J}$ & $\mathcal{F}$ & $\mathcal{J\&F}$ &
$\mathcal{J}$ & $\mathcal{F}$ & $\mathcal{J\&F}$ &
$\mathcal{J}$ & $\mathcal{F}$ & $\mathcal{J\&F}$ &
$\mathcal{S}$ \\
\midrule
AVSBench~\cite{zhou2022avs} & ECCV'22 & \multirow{4}{*}{AVS} & 23.2 & 51.1 & 37.2 & 32.4 & 54.7 & 43.5 & 27.8 & 52.9 & 40.3 & 0.208  \\

AVSegFormer~\cite{gao2024avsegformer} & AAAI'24 & & 34.5 & 47.0 & 40.2 & 36.1 & 50.1 & 43.1 & 34.8 & 48.6 & 41.7 & 0.171  \\

GAVS~\cite{wang2024prompting} & AAAI'24 & & 28.9 & 49.8 & 39.4 & 29.8 & 49.7 & 39.8 & 29.4 & 49.8 & 39.6 & 0.190\\
SAMA~\cite{liu2024annotation} & AAAI'24 & & 28.9 & 49.8 & 39.4 & 29.8 & 49.7 & 39.8 & 29.4 & 49.8 & 39.6 & 0.190 \\

\midrule
ReferFormer~\cite{wu2022language} & CVPR'22 & \multirow{2}{*}{Ref-VOS} & 31.3 & 50.1 & 40.7 & 30.4 & 48.8 & 39.6 & 30.9 & 49.5 & 40.2 & 0.176 \\
R2VOS~\cite{li2023robust} &  ICCV'23 &  & 25.0 & 41.0 & 33.0 & 27.9 & 49.8 & 38.9 & 26.5 & 45.4 & 35.9 & 0.183 \\
\midrule
EEMC~\cite{wang2024ref} & ECCV'24 & \multirow{4}{*}{Ref-AVS}  &  34.2 & 51.3 & 42.8 & 49.5 & 64.8 & 57.2 & 41.9 & 58.1 & 50.0 & \textbf{0.007} \\
Grounded-SAM2~\cite{ren2024grounded}& ArXiv'24 & & 28.5 & 39.9 & 34.2 & 59.8 & 68.1 & 63.9 & 44.2 & 54.0 & 49.1 & 0.277 \\
Crab~\cite{du2025crab} & CVPR'25 & & 40.5 & - & - & 45.6 & - & - & 43.1 & - & - & - \\

TSAM~\cite{radman2025tsam} & CVPR'25 &  & 43.4 & \underline{56.8} & \underline{50.1} & 54.6 & 66.4 & 60.5 & 49.0 & 61.6 & 55.3 & \underline{0.017} \\
SAM2-LOVE~\cite{wang2025sam2love} & CVPR'25 & & \underline{43.5} & 51.9 & 47.7 & \underline{66.5} & \underline{72.3} & \underline{69.4} & \underline{55.0} & \underline{62.1} & \underline{58.5} & 0.23 \\ 

\midrule
 \textbf{TGS-Agent (ours)} & - & Ref-AVS  & \textbf{49.5} & \textbf{60.4} & \textbf{54.9} & \textbf{73.2} & \textbf{80.6} & \textbf{76.9} & \textbf{61.3} & \textbf{70.5} & \textbf{65.9} & {0.035}\\
\bottomrule

\end{tabular}
}
\caption{Comparison with prior methods on Ref-AVSBench test set. $\mathcal{J\&F}$ is the average value of $\mathcal{J}$ and $\mathcal{F}$.}
\label{tab:ori_bench_results}
\end{table*}

\begin{table*}[t]
\centering
\resizebox{1\textwidth}{!}{
\begin{tabular}{l|l|ccc|ccc|ccc}
\toprule
& &
\multicolumn{3}{c|}{\textbf{Seen (S)} $\uparrow$} &
\multicolumn{3}{c|}{\textbf{Unseen (U)} $\uparrow$} &
\multicolumn{3}{c}{\textbf{Mix (S+U)} $\uparrow$}  \\
\textbf{Ref. Source} & \textbf{Method}   & 
$\mathcal{J}$ & $\mathcal{F}$ & $\mathcal{J\&F}$ &
$\mathcal{J}$ & $\mathcal{F}$ & $\mathcal{J\&F}$ &
$\mathcal{J}$ & $\mathcal{F}$ & $\mathcal{J\&F}$ \\
\midrule
\multirow{4}{*}{Ref-AVSBench} & Crab~\cite{du2025crab} & 20.7 & 33.8 & 27.2 & 39.4 & 54.9 & 47.2 & 30.1 & 44.4 & 37.2  \\
 & EEMC~\cite{wang2024ref}  & 26.4 & 40.9 & 33.7 & 39.7 & 55.9 & 47.8 & 33.1 & 48.4 & 40.7  \\
& Grounded-SAM2~\cite{ren2024grounded}  &  41.1 & 50.8 & 45.9 & 68.1 & 76.2 & 72.2 & 54.6 & 63.5 & 59.0 \\
&  \textbf{TGS-Agent (ours)}   & \textbf{48.5} & \textbf{58.4} & \textbf{53.4} & \textbf{71.8} & \textbf{80.4} & \textbf{76.1} & \textbf{60.1} & \textbf{69.4} & \textbf{64.8} \\
\midrule
\multirow{4}{*}{R\textsuperscript{2}-AVSBench} &Crab~\cite{du2025crab} &  19.6 & 32.1 &25.9 & 36.9 & 52.0 & 44.4 & 28.2 & 42.1 & 35.2 \\
 & EEMC~\cite{wang2024ref}   & 22.6 & 39.9 & 31.2 & 38.7 & 56.9 & 47.8 & 30.6 & 48.4 & 39.5 \\
& Grounded-SAM2~\cite{ren2024grounded}   & 21.4 & 30.1 & 25.7 & 44.9 & 53.4 & 49.1 & 33.2 & 41.7 & 37.4  \\
& \textbf{TGS-Agent (ours)}  &  \textbf{42.7} & \textbf{52.3} & \textbf{47.5} & \textbf{68.3} & \textbf{77.0} & \textbf{72.7} & \textbf{55.5 }& \textbf{64.7} & \textbf{60.1} \\

\bottomrule

\end{tabular}
}
\caption{{Evaluation on R\textsuperscript{2}-AVSBench.} We also report results on the same set of videos using original Ref-AVSBench references.}
\label{tab:our_bench_results}
\end{table*}

\subsection{Evaluation on Ref-AVSBench}
We compare our method with previous works on the standard Ref-AVSBench dataset.
Following the pioneering work~\cite{wang2024ref}, we report results of methods adapted from related AVS and Ref-VOS tasks, incorporating text and audio modalities, respectively.
As shown in Table~\ref{tab:ori_bench_results}, our TGS-Agent outperforms all these methods, including recent state-of-the-art approaches in the Ref-AVS domain.
For example, compared to SAM2-LOVE~\cite{wang2025sam2love}, our method surpasses it by 7.2\% and 7.5\% in $\mathcal{J}\&\mathcal{F}$ on the \textit{Seen} and \textit{Unseen} test sets, respectively.
The proposed Ref-Thinker, together with Grounding-DINO and SAM2, supports open-set vocabulary, making our method effective in both seen and unseen class scenarios.
Although both SAM2-LOVE and our method use the same SAM2 model, SAM2-LOVE has much higher $\mathcal{S}$ on the \textit{Null} set, indicating more false positive predictions.
Notably, 1) we test Grounded-SAM2~\cite{ren2024grounded} on Ref-AVS task, which directly uses the references as text prompts to generate segmentation masks. While this approach shows competitive performance on \textit{Unseen} set, it performs poorly on \textit{Seen} set.
In contrast, our TGS-Agent introduces Ref-Thinker, which interprets the reference alongside audio and visual signals to generate a more explicit object description, significantly improving the performance.
2) Crab~\cite{du2025crab} is also an MLLM, designed for unified audio-visual task learning, including Ref-AVS. Despite using the same LLM backbone, our method significantly outperforms Crab.
These improvements are attributed to our more effective agentic framework design and the instruction tuning strategy enhanced by explicit object-aware reasoning.

\subsection{Evaluation on R\textsuperscript{2}-AVSBench}
We evaluate our method on the proposed R\textsuperscript{2}-AVSBench.
We primarily compare against those open-sourced methods: Crab~\cite{du2025crab}, EEMC~\cite{wang2024ref}, and Grounded-SAM2~\cite{ren2024grounded}.
Notably, all these pretrained models are directly evaluated on R\textsuperscript{2}-AVSBench.
As shown in Table~\ref{tab:our_bench_results} (bottom four rows), our method consistently outperforms other competitors by a large margin. 
The upper part of the table reports results obtained on the same test videos but using the original Ref-AVSBench references.
We can observe a consistent performance drop for all models when evaluated with the transformed references from R\textsuperscript{2}-AVSBench, indicating the increased complexity.
In particular, Grounded-SAM2 suffers a substantial drop of around 20\% on both \textit{Seen} and \textit{Unseen} sets, highlighting its sensitivity to reference variation.
In contrast, benefiting from the object-aware reasoning mechanism, our method can accurately identify the referred object from multimodal cues, thereby maintaining superior performance and demonstrating stronger generalizability across diverse and challenging reference types.

\begin{table}[t]
\centering
\resizebox{\columnwidth}{!}{
\setlength{\tabcolsep}{2pt}
\begin{tabular}{l|ccc|ccc|c}
\toprule
& 
\multicolumn{3}{c|}{\textbf{Seen (S)} $\uparrow$} &
\multicolumn{3}{c|}{\textbf{Unseen (U)} $\uparrow$}  &
\textbf{Null} $\downarrow$ \\
\textbf{Setup}  &
$\mathcal{J}$ & $\mathcal{F}$ & $\mathcal{J\&F}$ &
$\mathcal{J}$ & $\mathcal{F}$ & $\mathcal{J\&F}$ &
$\mathcal{S}$ \\
\midrule
Reference ($\bm{R}$)  & 28.5 & 39.9 & 34.2 & 59.8 & 68.1 & 63.9  & 0.277  \\
F-object ($\bm{T_f}$)  & 43.9 & 54.1 & 49.0 & 69.9 & 77.4 & 73.7 &  0.043 \\
\textbf{S-object  ($\bm{T_s}$}) & \textbf{49.5} & \textbf{60.4} & \textbf{54.9} & \textbf{73.2} & \textbf{80.6} & \textbf{76.9}  & \textbf{0.035}\\
\bottomrule

\end{tabular}
}
\caption{Ablation on object text used as prompts.}
\label{tab:ablation_f_s_objects}
\end{table}

\subsection{Ablation Study and Qualitative Results}\label{sec:ablations}
In this section, the experiments are conducted on the Ref-AVSBench dataset unless otherwise specified.

\begin{figure}[!t]
\centering
 \includegraphics[width=1\columnwidth]{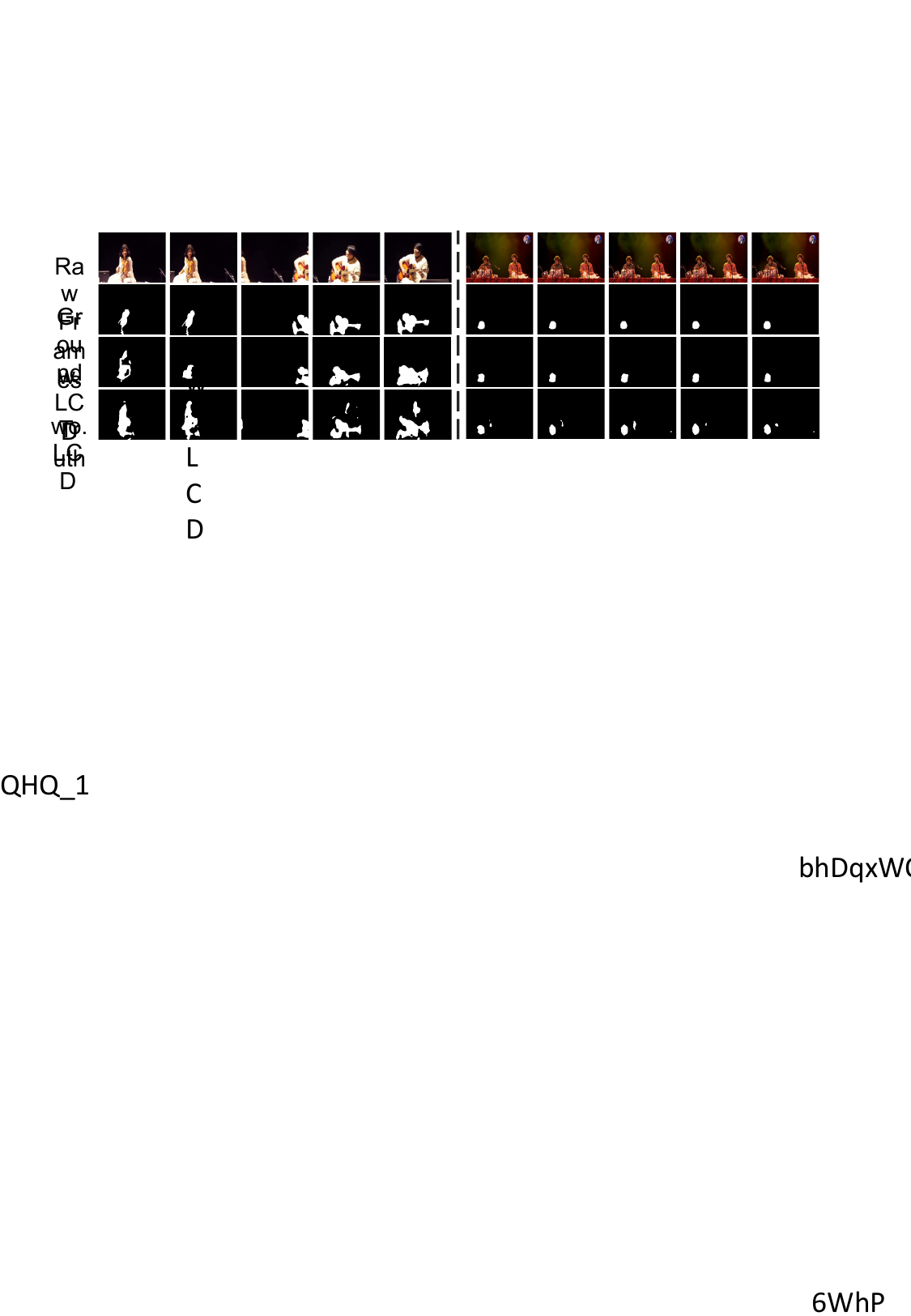}
\caption{Visualization results of using different object descriptions as detection prompts. The fine-grained phase $\bm{T_f}$ may be more helpful in video scenes containing multiple instances of the same category (\textit{i.e.}, \textit{guitar} in this example).
 }
 \label{Fig:f_better_than_s}
\end{figure}

\noindent\textbf{Impact of Object Description Types.}
Our method designs Ref-Thinker to generate two types of object descriptions, \textit{i.e.}, a fine-grained version $\bm{T_f}$ containing more details (f\_object), and a simplified version $\bm{T_s}$ consisting only of the object name (s\_object).
These are used as prompts in the \textit{Ground} phase for bounding box generation.
As shown in Table~\ref{tab:ablation_f_s_objects}, compared to using original reference expressions, both $\bm{T_f}$ and $\bm{T_s}$ significantly improve the performance, highlighting the benefits of explicit object-aware reasoning.
Interestingly, the simplified version (\textit{i.e.}, object category) yields better performance.
We hypothesize that this is related to the Grounding-DINO detector's preference, which may be more friendly to short and specific prompts.
However, in some cases, the fine-grained object phase can be more effective, especially when there are multiple similar/same object instances.
As shown in Fig.~\ref{Fig:f_better_than_s}, two guitars are played by different individuals.
Using only the simplified object class \textit{guitar} ($\bm{T_s}$) as a prompt, the object detector is confused when recognizing the correct guitar at instance level.
In contrast, the fine-grained phase ($\bm{T_f}$) provides richer cues, specifying that the referred guitar is \textit{played by the man}, which helps to improve the detection and segmentation results.
Nevertheless, even using $\bm{T}_f$, our method surpasses the prior state-of-the-art SAM2-LOVE~\cite{wang2025sam2love} (Table~\ref{tab:ori_bench_results}), demonstrating the superiority of our method.

\begin{figure}[!t]
\centering
 \includegraphics[width=1\columnwidth]{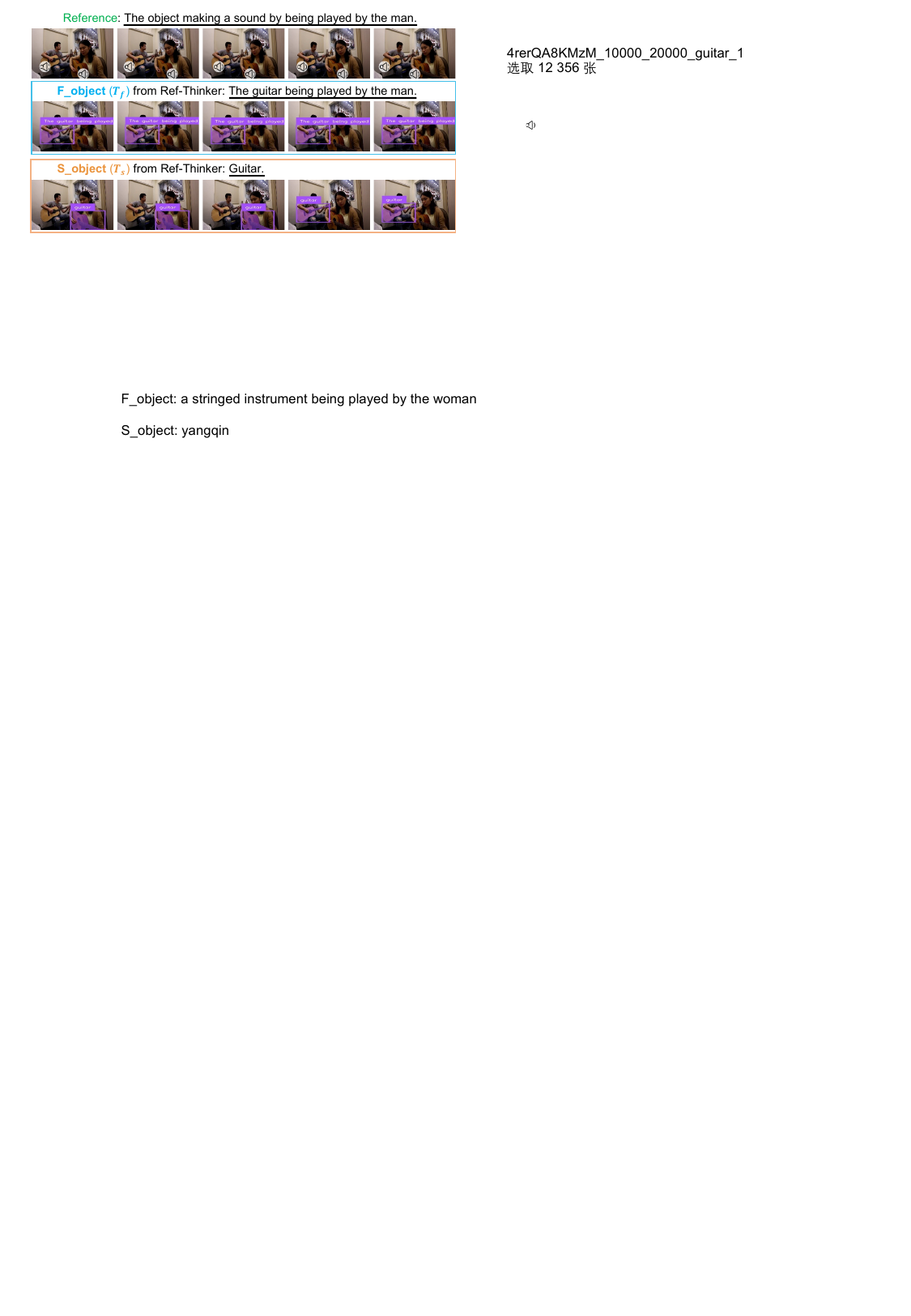}
\caption{Parameter study on the $\tau_\text{bbox}$ and $\tau_\text{text}$.
 }
 \label{Fig:thre_study}
\end{figure}

\noindent\textbf{Impact of Threshold Parameters.}
The \textit{Ground} phase of our TGS-Agent involves two hyperparameters, namely the bounding box threshold $\tau_{\text{bbox}}$ and the text similarity threshold $\tau_{\text{text}}$, which are used to control the selection of object bounding boxes.
We investigate their influence and present the results in Fig.~\ref{Fig:thre_study}.
The model performance shows slight variation across different values of $\tau_{\text{bbox}}$, while remaining stable with respect to $\tau_{\text{text}}$.
We set $\tau_{\text{bbox}} = 0.1$ and $\tau_{\text{text}} = 0.25$ as the default configuration in our experiments.

\begin{figure}[!t]
\centering
 \includegraphics[width=1\columnwidth]{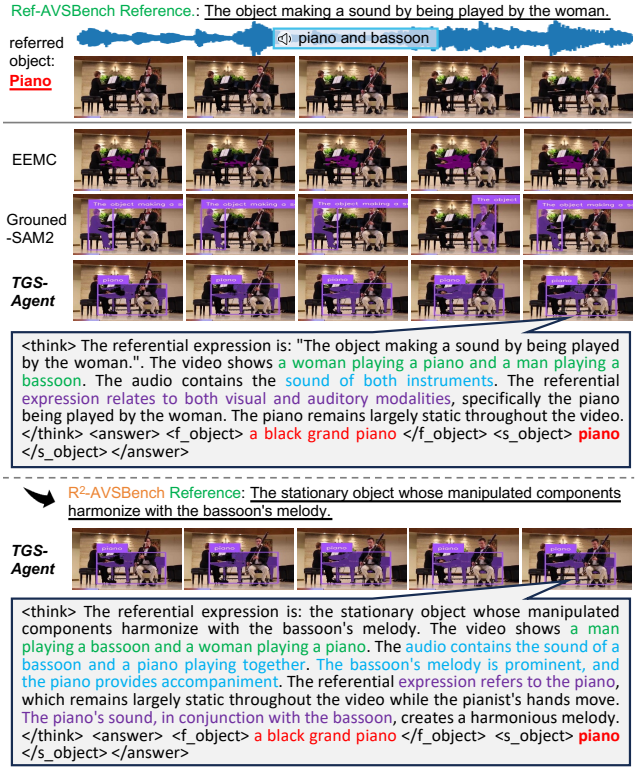}
\caption{Qualitative example of the Ref-AVS task.}
 \label{Fig:method_seg_results}
\end{figure}

\noindent\textbf{Qualitative Analysis.}
Fig.~\ref{Fig:method_seg_results} presents some qualitative results.
In the video example, a man is playing a bassoon while a woman is playing a piano.
The referred object is the \textit{piano}.
Prior method EEMC~\cite{wang2024ref} identifies the piano but only segments part of it.
Grounded-SAM2, which directly uses the reference expression for prediction, incorrectly segments another object, \textit{e.g., the woman}, mentioned in the reference text.
In contrast, our TGS-Agent method accurately identifies the referred object and achieves satisfactory segmentation.
This improvement stems from our Ref-Thinker, which thoroughly analyzes the reference alongside the audio and visual content, and then explicitly outputs the target object for downstream tool invocation.
The corresponding reasoning process is visualized in the figure.
In addition, we provide results of our model on more complex reference from our R\textsuperscript{2}-AVSBench.
As illustrated, our Ref-Thinker can adapt its reasoning to new reference and still successfully identify the referred object.
We provide more qualitative results and some failure case studies in Appendix D\&E.

\section{Conclusion}
We propose TGS-Agent, addressing the Ref-AVS task through an agentic \textit{Think-Ground-Segment} paradigm.
At its core is Ref-Thinker, a reasoning-enhanced MLLM that transforms multimodal inputs (reference text and audio-visual streams) into explicit descriptions of the referred object, which better prompts subsequent grounding and segmentation processes.
To further benchmark model robustness, we introduce R\textsuperscript{2}-AVSBench, a new evaluation set featuring linguistically diverse and reasoning-intensive references.
R\textsuperscript{2}-AVSBench can be used for assessing a model's cross-reference generalization and may benefit the broader community as a challenging testbed. 
Our TGS-Agent operates without requiring pixel-level supervision, offering greater explainability.
Despite its simplicity, it significantly outperforms prior models.
Overall, our work highlights the importance of \textit{explicit reference understanding} in Ref-AVS.
Future directions include integrating our Ref-Thinker with SAM2 tuning-based methods to further enhance the performance.

\bibliography{aaai2026}

\clearpage

\appendix

In this appendix, we provide detailed prompts used for constructing both the instruction tuning set (Sec.~\ref{sec:prompt_instruction_tuning_set}) and the R\textsuperscript{2}-AVSBench (Sec.~\ref{sec:prompt_R2AVSBench}).
Additionally, we present more visualization examples of our R\textsuperscript{2}-AVSBench (Sec.~\ref{sec:supp_examples_R2AVSBench}), more qualitative comparison results (Sec.~\ref{sec:supp_qualitative_results}), and the failure case analysis of our method (Sec.~\ref{sec:supp_failure_case_study}).

\section{Prompt used for Instruction Tuning Set}\label{sec:prompt_instruction_tuning_set}
The prompt is provided in Fig.~\ref{Fig:prompt_tuning_set}.

\section{Prompt used for Building R\textsuperscript{2}-AVSBench}\label{sec:prompt_R2AVSBench}
The prompt is provided in Fig.~\ref{Fig:prompt_our_bench}.

\section{Reference Examples of R\textsuperscript{2}-AVSBench}\label{sec:supp_examples_R2AVSBench}
\begin{figure*}[!t]
\centering
 \includegraphics[width=1\textwidth]{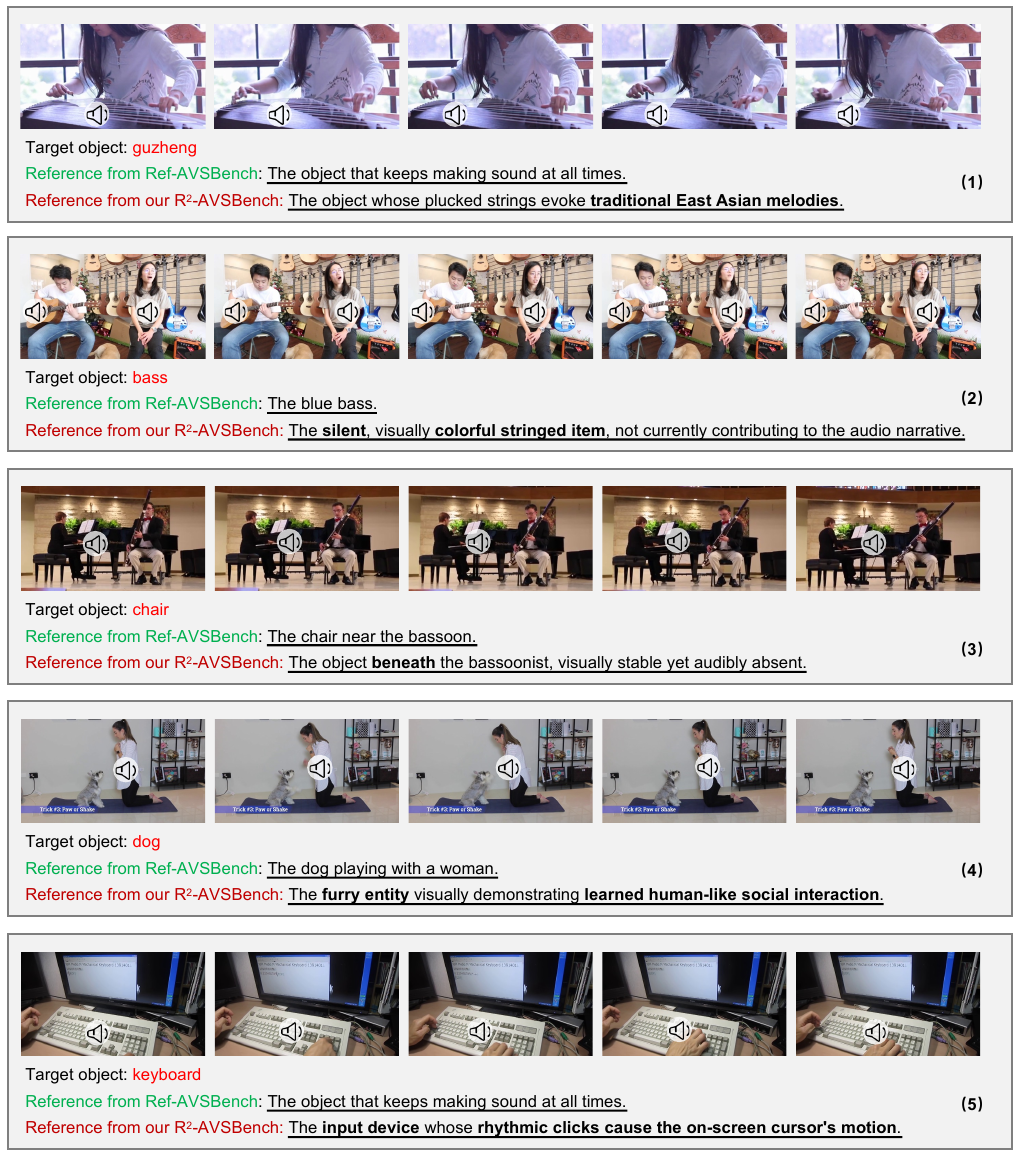}
\caption{Examples of reference expressions from our R\textsuperscript{2}-AVSBench. Compared to the original references in Ref-AVSBench~\cite{wang2024ref}, ours are more linguistically complex and reasoning-intensive.
 }
 \label{Fig:our_bench_examples}
\end{figure*}

In Fig.~\ref{Fig:our_bench_examples}, we present several example references from our proposed R\textsuperscript{2}-AVSBench evaluation set.
Compared to the references in the widely-used Ref-AVSBench~\cite{wang2024ref} dataset, our references exhibit greater linguistic complexity, diversity, and reasoning demands.
For instance, in example (1), the target object is \textit{guzheng}. The phrase \textit{“traditional East Asian melodies”} in our reference requires reasoning over cultural and historical knowledge.
In examples (2), (3), and (4), the original Ref-AVSBench references explicitly mention the target objects, whereas our transformed references necessitate inference based on color and material (2), spatial relationships (3), or object behavior (4).
The final example (5) illustrates a case where reasoning about the object’s function is required.

Overall, these examples highlight the high quality of our transformed references: they target the same object as the original but in a more linguistically diverse and reasoning-intensive manner.
This makes R\textsuperscript{2}-AVSBench a valuable benchmark for evaluating a model’s robustness and generalization across varied reference types.
We will release R\textsuperscript{2}-AVSBench to the community.

\section{More Qualitative Comparison Results}\label{sec:supp_qualitative_results}
\begin{figure*}[!t]
\centering
 \includegraphics[width=1\textwidth]{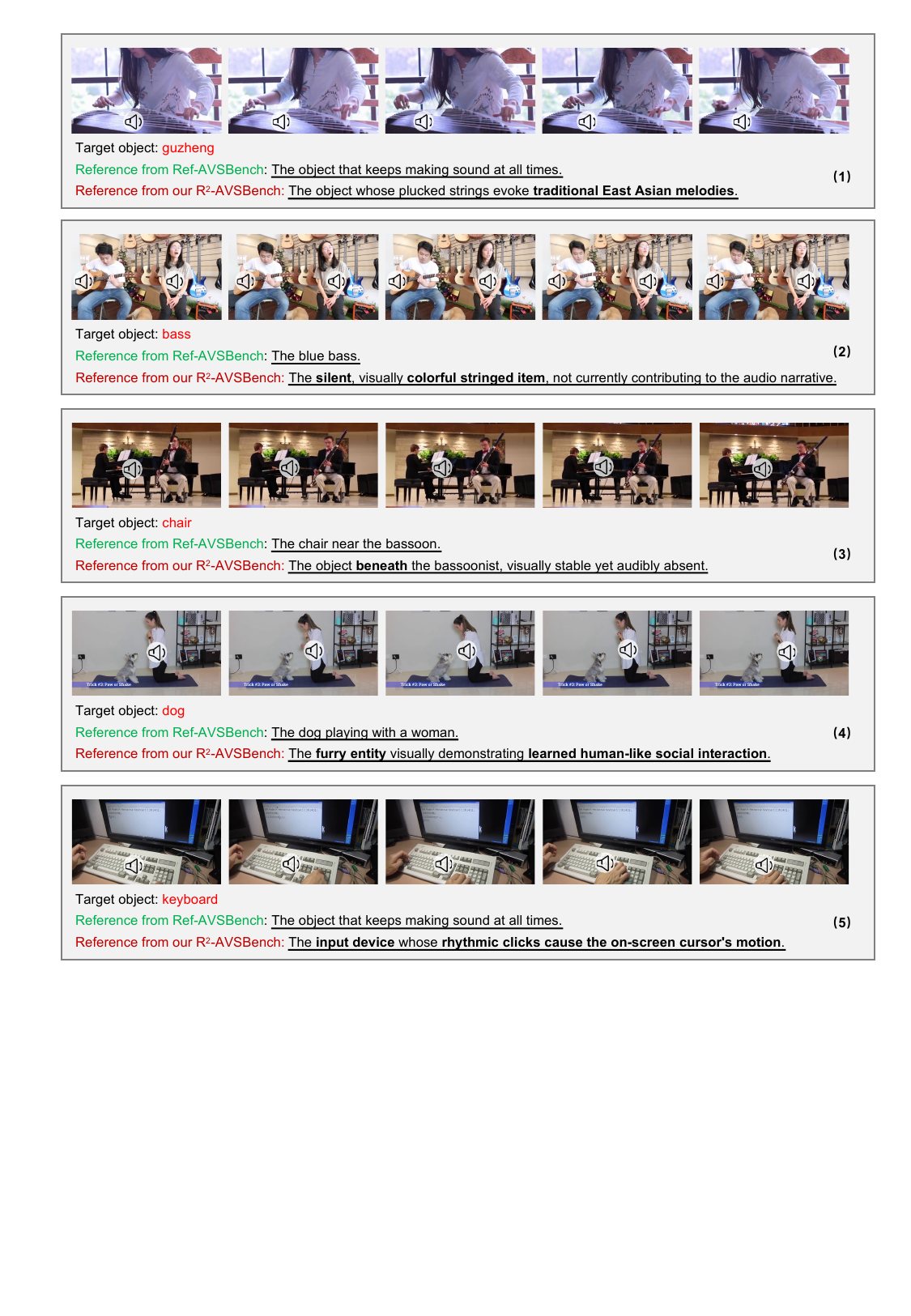}
\caption{Comparison results on the Ref-AVSBench dataset: \textit{Seen} test set samples.}
 \label{Fig:supp_quali_ori_bench_seen_examples}
\end{figure*}

\begin{figure*}[!t]
\centering
 \includegraphics[width=1\textwidth]{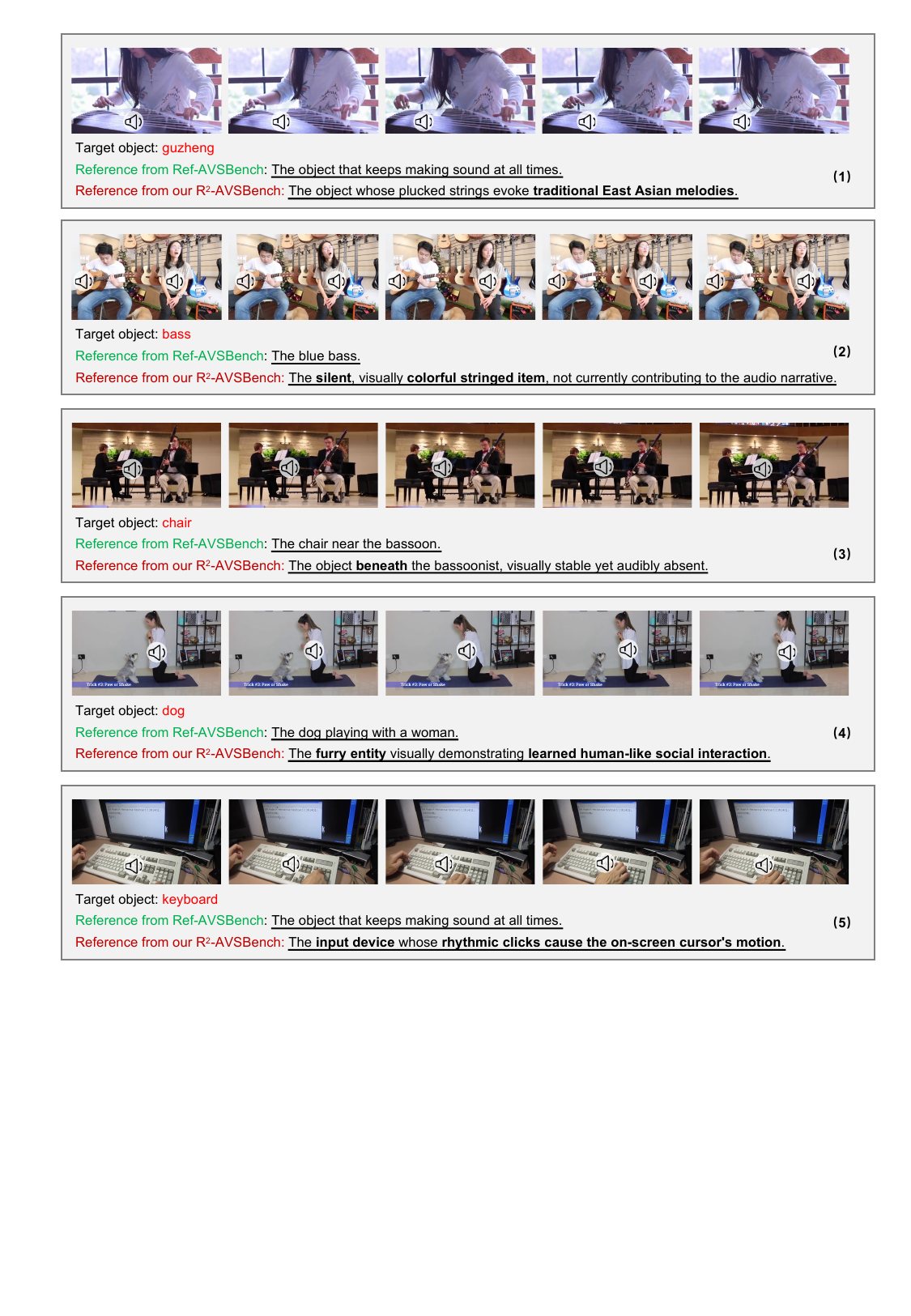}
\caption{Comparison results on the Ref-AVSBench dataset: \textit{Unseen} test set samples.}
 \label{Fig:supp_quali_ori_bench_unseen_examples}
\end{figure*}

\begin{figure*}[!t]
\centering
 \includegraphics[width=1\textwidth]{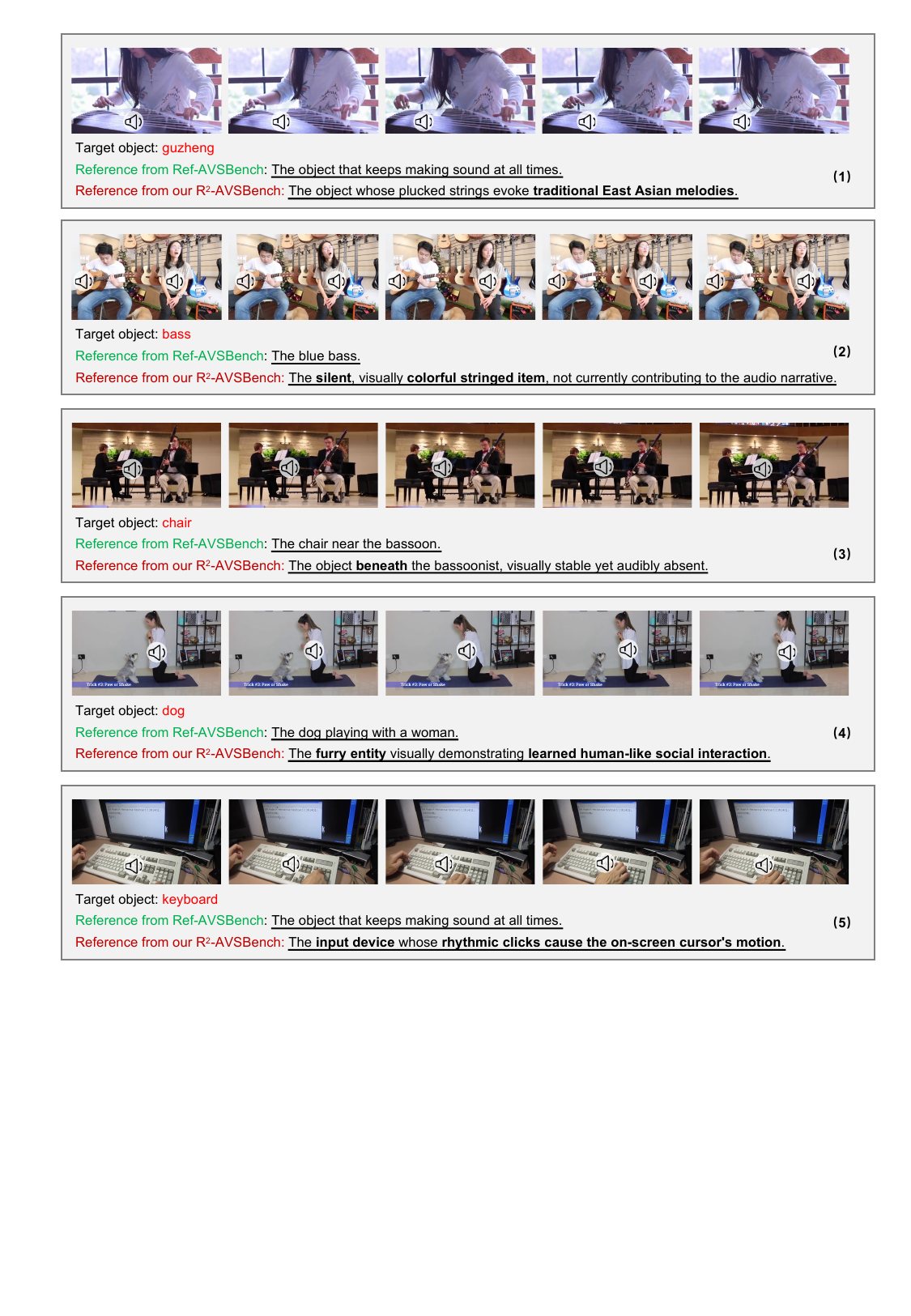}
\caption{Comparison results on the R\textsuperscript{2}-AVSBench dataset: \textit{Seen} test set samples.}
 \label{Fig:supp_quali_our_bench_seen_examples}
\end{figure*}

\begin{figure*}[!t]
\centering
 \includegraphics[width=1\textwidth]{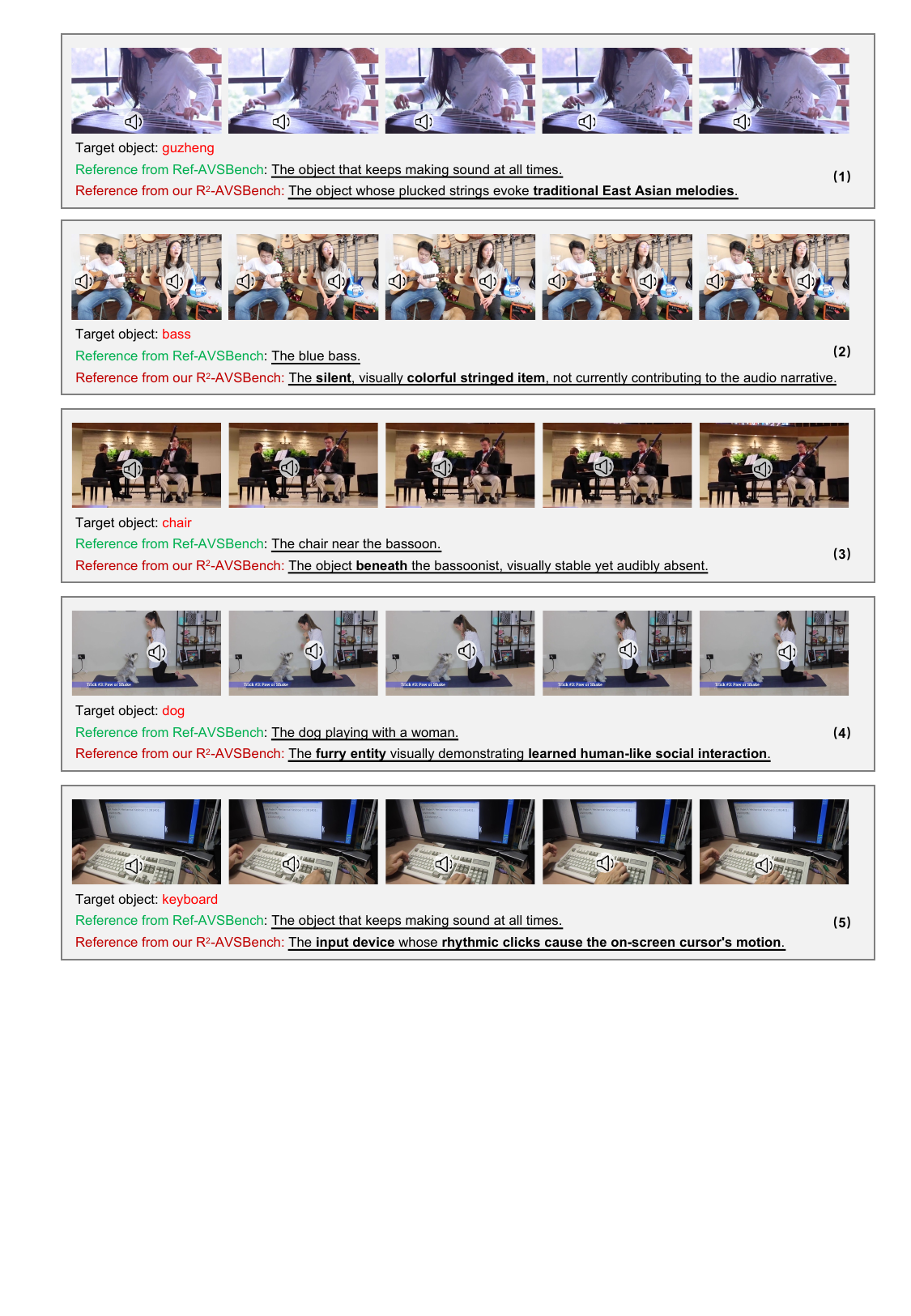}
\caption{Comparison results on the R\textsuperscript{2}-AVSBench dataset: \textit{Unseen} test set samples.}
 \label{Fig:supp_quali_our_bench_unseen_examples}
\end{figure*}

We provide additional qualitative results, including comparisons with prior methods, EEMC~\cite{wang2024ref} and Grounded-SAM2~\cite{ren2024grounded}. Specifically, we present results on the \textit{Seen} (Fig.\ref{Fig:supp_quali_ori_bench_seen_examples}) and \textit{Unseen} (Fig.\ref{Fig:supp_quali_ori_bench_unseen_examples}) test sets of Ref-AVSBench, as well as the \textit{Seen} (Fig.\ref{Fig:supp_quali_our_bench_seen_examples}) and \textit{Unseen} (Fig.\ref{Fig:supp_quali_our_bench_unseen_examples}) test sets of our proposed R\textsuperscript{2}-AVSBench.

These visualization results demonstrate that our TGS-Agent consistently produces more accurate segmentation results aligned with the referred objects. For instance, EEMC~\cite{wang2024ref} frequently over-segments irrelevant objects—such as the \textit{left guitar} in Fig.\ref{Fig:supp_quali_ori_bench_seen_examples}a, the \textit{mouse pad} in Fig.\ref{Fig:supp_quali_ori_bench_seen_examples}b, the \textit{man} in Fig.\ref{Fig:supp_quali_ori_bench_unseen_examples}a, and the \textit{tuba} in Fig.\ref{Fig:supp_quali_ori_bench_unseen_examples}b—and fails to capture the correct target objects in more challenging cases from R\textsuperscript{2}-AVSBench, such as the \textit{cello} in Fig.\ref{Fig:supp_quali_our_bench_seen_examples}b and the \textit{baby} in Fig.\ref{Fig:supp_quali_our_bench_unseen_examples}b.
Similarly, Grounded-SAM2, which directly relies on reference texts for segmentation, struggles when multiple objects are mentioned in the reference expression. This leads to incorrect predictions in cases like the \textit{chair} in Fig.\ref{Fig:supp_quali_ori_bench_seen_examples}a, the \textit{flute} in Fig.\ref{Fig:supp_quali_ori_bench_unseen_examples}b, and the \textit{cello} in Fig.~\ref{Fig:supp_quali_our_bench_seen_examples}b.
In contrast, our method produces more precise results, primarily due to the effectiveness of the proposed Ref-Thinker, which explicitly interprets the reference expression in conjunction with the audio and visual cues before guiding segmentation.

\textbf{Analysis on the reasoning accuracy of the predicted object.} For each sample shown in the figures, we also visualize the reasoning process of Ref-Thinker, which successfully analyzes the reference, visual content (highlighted in green), audio signal (in blue), and modality-specific focus (in purple), ultimately producing an accurate and interpretable object description.
Beyond qualitative analysis, we also provide a quantitative evaluation of the accuracy of the simplified object descriptions (`s\_object', denoted as $\bm{T_s}$ in the main paper).
We compare $\bm{T_s}$ with the ground truth object categories.
Our results show that Ref-Thinker precisely matches the annotated object category for 55.1\% of the \textit{Seen} test set and 49.3\% of the \textit{Unseen} test set on the Ref-AVSBench.
It is worth noting that the output of Ref-Thinker is open-vocabulary, and thus may not exactly match the fixed annotations of Ref-AVSBench, even when semantically correct.
To further assess semantic alignment, we compute CLIPScore~\cite{hessel2021clipscore} between the predicted descriptions and the ground-truth categories. 
We observe that the majority of CLIPScore values fall within the range of $[0.8, 1]$, indicating strong semantic similarity.
For instance, as shown in Fig.~\ref{Fig:supp_quali_our_bench_unseen_examples}b, although the annotated label of the referred object is \textit{baby}, our Ref-Thinker outputs \textit{child}, which still serves as a reliable and distinctive cue for guiding the segmentation process.

\section{Failure Case Analysis}\label{sec:supp_failure_case_study}

\begin{figure*}
\centering
 \includegraphics[width=1\textwidth]{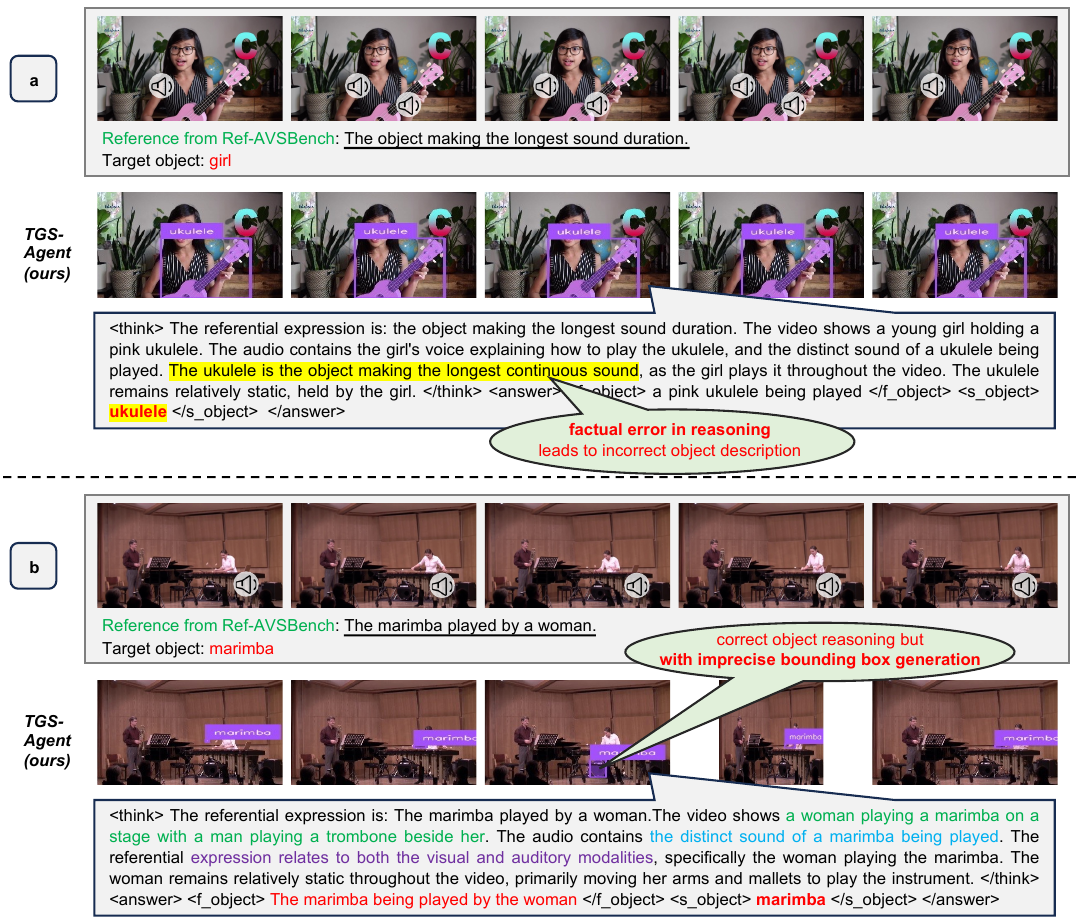}
\caption{Failure cases analysis. Samples are from the Ref-AVSBench dataset.}
 \label{Fig:supp_failure_cases}
\end{figure*}

While our TGS-Agent demonstrates strong performance through both qualitative and quantitative evaluations, it may still fail in certain challenging scenarios. We provide a further analysis below:

\textbf{1) Factual errors at the \textit{Think} stage.}
As illustrated in Fig.~\ref{Fig:supp_failure_cases}a, both the ukulele and the girl produce sound, and the reference specifies the object \textit{making the longest sound duration}. Although Ref-Thinker successfully identifies the presence of both the girl's voice and the ukulele sound, it incorrectly concludes that the \textit{ukulele} has the longest continuous sound. This failure highlights a limitation in fine-grained temporal reasoning over mixed audio streams, which remains a challenging problem. It also motivates future research on improving temporal audio understanding and better balancing multimodal content analysis.

\textbf{2) Imprecise bounding boxes despite correct reasoning.}
In Fig.~\ref{Fig:supp_failure_cases}b, Ref-Thinker correctly interprets the reference and identifies the target object as \textit{marimba}. However, during the \textit{Ground} phase, the object detector (Grounding-DINO) fails to generate accurate bounding boxes. This may stem from limited pre-training data involving rare objects like the \textit{marimba}. Nonetheless, such limitations can potentially be addressed by incorporating more diverse and representative training data or leveraging advanced object detection techniques.
Empirically, we find that the subsequent \textit{Segment} phase, powered by SAM2, performs well when provided with precise bounding boxes.


\begin{figure*}[t]
\centering
\begin{tcolorbox}[
  colback=gray!10,
  colframe=black,
  arc=3mm,
  boxrule=0.8pt,
  width=0.95\textwidth,
  enlarge left by=0mm,
  enlarge right by=0mm
]
You are an expert multimodal AI trainer, assisting in creating high-quality CoT-style instruction data for training a multimodal model. 

Your task is to generate a $<$think$>$ reasoning chain and an $<$answer$>$ output for each example based on the following inputs:

- ref: A referential expression describing a single object. The object may be referenced by visual features (video frames), auditory features (audio), or both.\\
- video frames: 10 raw video frames (images), each resized to 224x224 resolution.\\
- audio file: A corresponding .wav file representing the audio track of the video.\\

\textbf{Guidelines for $<$think$>$:}\\
1. In $<$think$>$, you must:

- Start your $<$think$>$ reasoning with: The referential expression is: "$<$ref$>$". Do not alter the given ref in any way.

- First provide a brief description of the overall visual and audio context. If the audio has been explicitly indicated as silent, state 'The audio is silent.' directly. Otherwise, analyze the provided audio file. If the audio is silent or irrelevant to the visual content, explicitly state this. Then, shift your reasoning focus on analyzing the object referred by the ref. Your reasoning may integrate visual, auditory, and textual information.

- Analyze the semantics of the ref: object attributes (e.g., appearance, action, sound, position, temporal order). Whether the ref is related to visual, audio, or both modalities. If the audio has been explicitly indicated as silent, do not describe any sounds from it. Otherwise, if the audio is truly silent or contains only background noise, acknowledge this clearly.

- You may analyze the motion pattern of the object across frames, determining if the object is mostly static, moving slightly, or showing significant movement across frames.

- Keep reasoning concise or approximately 50-60 words, focused, and based only on the given information. Avoid overthinking or making assumptions beyond the provided data.\\

\textbf{Guidelines for $<$answer$>$:}\\
2. In $<$answer$>$, output must follow this strict format:\\
\hspace*{3mm}$<$f\_object$>$ A fine-grained description of the object (appearance, location, attributes, actions) $<$/f\_object$>$\\
\hspace*{3mm}$<$s\_object$>$ The simplified category of the object (e.g., guitar, dog, car) $<$/s\_object$>$

- Keep the $<$f\_object$>$ description concise (6-10 words). If the object is unique in the scene, a slightly more detailed description than $<$s\_object$>$ is sufficient. If there are multiple instances of the same category, the $<$f\_object$>$ must include distinguishing features such as position, color, attributes, or actions.

- $<$f\_object$>$ and $<$s\_object$>$ must describe the same object referred by the ref.

- If the object is not present, output:\\
  \hspace*{3mm}$<$f\_object$>$ null $<$/f\_object$>$\\
  \hspace*{3mm}$<$s\_object$>$ null $<$/s\_object$>$
  \\

Strictly follow the tag format. Each tag ($<$f\_object$>$, $<$s\_object$>$, $<$think$>$, $<$answer$>$, etc.) must appear on its own line. Do not merge tags, omit line breaks, or add extra whitespace.

Expected Output Example:\\
$<$think$>$\\
The referential expression is: "a person holding a guitar". The video shows a person actively playing a guitar, moving slightly across frames. The audio contains clear guitar strumming sounds. The referential expression primarily relates to the visual and auditory presence of the person and the guitar.\\ 
$<$/think$>$\\
$<$answer$>$\\
\hspace*{3mm}$<$f\_object$>$\\
\hspace*{6mm}a person holding a guitar, shifting position from left to right\\
\hspace*{3mm}$<$/f\_object$>$\\
\hspace*{3mm}$<$s\_object$>$ \\
\hspace*{6mm}person \\
\hspace*{3mm}$<$/s\_object$>$\\
$<$/answer$>$\\

\end{tcolorbox}
\caption{Prompt used for constructing Ref-Thinker instruction tuning set.}\label{Fig:prompt_tuning_set}
\end{figure*}

\begin{figure*}[t]
\centering
\begin{tcolorbox}[
  colback=gray!10,
  colframe=black,
  arc=3mm,
  boxrule=0.8pt,
  width=0.95\textwidth,
  enlarge left by=0mm,
  enlarge right by=0mm
]
You are an expert in multimodal dataset construction. Your core task is to {generate a novel, highly challenging, and reasoning-rich referring expression (ref) for a specified target object within a given video context (identified by UID and pixel mask). } This new ref must precisely identify the target object, but its recognition process {ABSOLUTELY MUST require advanced, complex reasoning from an MLLM, going far beyond simple attribute matching or direct translations.}
\\

You will be provided with:\\
- A \texttt{uid}: \texttt{\{uid\}}, with its associated {target object} name: {"\{target\_object\_name\}"}.\\
- \texttt{\{mask\_mention\}}\\
- Actual video frames (10 frames) and an audio track. {It is PARAMOUNT that you comprehensively analyze and fuse both visual and audio information to understand the {target object}'s unique characteristics, behaviors, and interactions. The generated \texttt{complex\_ref} MUST reflect this deep multimodal understanding, not just a single modality.}\\

For this current \texttt{uid} (\texttt{\{uid\}}), your goal is:\\
1. {To create a completely new \texttt{complex\_ref} for the {target object} ("\{target\_object\_name\}").} This \texttt{complex\_ref} must {unambiguously and accurately refer to the exact target object} uniquely identified by the \texttt{uid} and its associated pixel mask in the video.\\

2. Your \texttt{complex\_ref} must be {ABSOLUTELY CONCISE (STRICTLY 5-15 WORDS, NO EXCEPTIONS)} and {strictly} incorporate one or more of these challenging reasoning types. {Crucially: avoid any simple, direct descriptions. If a direct description (like "red car" or "loud dog barking") can identify the object, your \texttt{complex\_ref} is considered INSUFFICIENT and will be REJECTED, regardless of length.}\\
    * {External Knowledge Reasoning:} Describe the target by referencing {general knowledge, cultural context, typical characteristics, or behavioral patterns} of the object from outside the video. This requires the MLLM to associate and reason with its stored knowledge of the external world, combined with visual and auditory cues from the video.

    * {Multimodal Comparative Reasoning:} Describe the target by comparing its visual, auditory, or behavioral aspects to {other objects in the scene} or {external knowledge}, focusing on non-obvious distinctions that require deep inference and cross-modal analysis.

    * {Multimodal Abstract/Functional/Role-based Reasoning:} Refer to the object by its implicit purpose, role, or an inferred characteristic based on its {visual state, auditory output, or complex interaction with the environment/agents.}

    * {Multimodal Complex Temporal/Causal Reasoning:} Link the object to non-obvious sequences, causes, or effects across time, interpreting events or changes in relation to the masked object's {visual state, auditory output, or their interplay.}\\


3. {ABSOLUTELY CRITICAL RESTRICTION: AVOID DIRECT ATTRIBUTES AND SIMPLE MULTIMODAL COMBINATIONS.}\\
    * {DO NOT} use explicit, easily identifiable attributes like {color} ("red car", "blue bird"), {simple shape} ("round ball", "square box"), {obvious size} ("large building", "tiny bug"), {direct, dominant sounds} ("loud alarm", "siren", "dog barking", "engine roaring"), or {simple actions} ("running man", "flying bird").\\
    * Your \texttt{complex\_ref} must NOT be a mere synonym replacement or simple rephrasing. It must be a truly novel description, forcing the MLLM to use contextual, comparative, functional, or temporal reasoning by {deeply integrating visual and audio cues based on the {target object} and its {pixel mask}}.\\
    * The challenge MUST originate from the required \textit{multimodal, complex reasoning}, NOT just more descriptive words from a single modality, nor simple combinations of direct attributes from multiple modalities.\\

4. {Output Format:} Return a JSON object containing the result for the current \texttt{uid}.\\



{5. Examples (New Style of HIGHLY CHALLENGING, MULTIMODAL, Reasoning-focused Refs - Concise and Complex):}\\
- {Target Object:} "hair-dryer" (Pixel mask shows it in hand, pointed at hair, generating sound and hot air movement.)\\
    {Reasoning-intensive Ref:} "{The handheld device creating localized heat and continuous ambient noise.}" (11 words)\\
- {Target Object:} "emergency-car" (Pixel mask shows an emergency vehicle, with flashing lights, but faint sound.)\\
    {Reasoning-intensive Ref:} "{A mobile entity visually indicating rescue intent, yet lacking its customary audible warning.}" (16 words)\\


\end{tcolorbox}
\caption{Prompt used for constructing R\textsuperscript{2}-AVSBench.}\label{Fig:prompt_our_bench}
\end{figure*}

\end{document}